\documentclass[twocolumn,aps,showpacs,prb,tightenlines,amssymb]{revtex4}
\usepackage{epsfig}
\usepackage{dcolumn}% Align table columns on decimal point
\usepackage{bm}% bold math

\begin{document}

\small
\title{Intense Terahertz Laser Fields Induced and Manipulated Pseudospin Polarization in Graphene}

\author{J. Zhou}
\email{zhouj3@mail.ustc.edu.cn}% 
\affiliation{Department of Physics, University of Science \& Technology of China, Hefei, Anhui, 230026, China}
\date{\today}

\begin{abstract}
We investigate the pseudospin-dependent density-energy relation (whose differential with respect to energy is
density of states) of a monolayer graphene under intense terahertz
laser field by exactly solving the time-dependent Schr\"{o}dinger equation with help of Floquet's theorem. 
We find that psedospin polarization can be induced by a circular polarized terahertz laser
field. The psedospin polarization in K and K$^{\prime}$ valleys can be exactly opposite sign when the electron
densities in these two calleys are equivalent. 
Further more, we find that the psedospin polarization can be manipulated by the strength,
frequency, and especially the polarization orientation of the field. 
\end{abstract}
\maketitle

After the success of isolating the one carbon atom thick film
graphene,\cite{novoselov1,novoselov2} which is the thinnest material throughout 
scientific history, scientists are exploiting this new field\cite{novoselov3,geim1,geim2,zhang,zhou,trau} both
theoretically and experimentally although it has been first studied more than 50 years ago.\cite{wallace} 
Such a single layer graphene system contains two distinguishable carbon atoms A and
B in each unit cell (A and B sublattice). The bonding $\pi$ and antibonding $\pi^{\ast}$ bands touch with each other near the Fermi
energy at two independent, degenerate and chiral symmetric Dirac points K and K$^{\prime}$ of the first Brillouin zone. 
Near these two points, the unusual dispersion which is approximately linear can form conically shaped valleys. 
It is intresting that the electron behavior in these valleys are govern by
relativistic Dirac equation and the charge carriers can be regarded as massless Dirac
Fermions\cite{novoselov3,zhang} with a effective speed of light $v_{F}\sim 1/400 c$.
However, one of the most important potential applications of graphene is the graphene-based electronic devices. For
example, the valley degree of freedom,\cite{rycerz} sublattice degree of freedom,\cite{tan} and layer index
in graphene bilayers\cite{min} are suggested to be used in the same way as the electron spin is
used in spintronics devices.\cite{wolf,prinz} 
It have been proposed recently that intense terahertz (THz) field is one possible tool for contralinng and
manipulating electron spin.\cite{johnsen,cheng,jiang1,jiang2} In this paper, we propose a sublattice
polarization generator by using circular polarized terahertz laser.   

We describe the sublattice degree of freedom as a {\em ``pseudospin''}.\cite{geim2}
The effective-mass Hamiltonian of an ideal infinite graphene can be written as:   
\begin{equation}
H_{0}({\bf k})=\hbar v_{F}[\sigma_{x}\tau k_{x}+\sigma_{y}k_{y}] \ ,
\end{equation}
where the Fermi velocity $v_{F}\approx 8\times 10^5 m/s$, $\tau=+ (-)$ represent the states in K (K$^{\prime}$)
valley. Here ${\bf \sigma}$ is the Pauli matrix of pseudospin, $\sigma_{z}=\pm 1$ 
denoting states on A (B) sublattice {\em i. e.} ``up'' (``down'') state of pseudospin.
We describe a circular polarized THz laser field ${\bf E}(t)=\frac{E_{0}}{\sqrt{2}}(\hat{x}\cos{\Omega
  t}+\hat{y}\sin{\Omega t})$ with period $T_{0}=\frac{2\pi}{\Omega}$, the vector and scalar potentials can be
chosen as ${\bf A}(t)=-\frac{E_{0}}{\sqrt{2}\Omega}(\hat{x}\sin{\Omega t}-\hat{y}\cos{\Omega t})$ and $\phi=0$
under Coulomb gauge.
With the gauge covariant derivative operator $P=-i\hbar \nabla-e{\bf A}(t)$, the time dependent Hamiltonian can be written as:
\begin{equation}
H({\bf k},t)=\hbar v_{F}\Big[\sigma_{x}\tau \Big(k_{x}+\frac{eE_{0}}{\sqrt{2}\hbar\Omega}\sin{\Omega
    t}\Big)+\sigma_{y}\Big(k_{y}-\frac{eE_{0}}{\sqrt{2}\hbar\Omega}\cos{\Omega t}\Big)\Big] \ .
\end{equation}
In order to understand the effect of circular polarized field clearer, we also need to know the effect of
linear polarized field for comparison. 
We have ${\bf E_{\tiny Linear}}(t)=E_{0}\hat{x}\cos{\Omega t}$, ${\bf A}(t)=-\frac{E_{0}}{\Omega}\hat{x}\sin{\Omega t}$, the
according Hamiltonian is:
\begin{equation}
H_{\mbox{\tiny Linear}}({\bf k},t)=\hbar v_{F}\Big[\sigma_{x}\tau \Big(k_{x}+\frac{eE_{0}}{\hbar\Omega}\sin{\Omega
    t}\Big)+\sigma_{y}k_{y}\Big] \ .
\end{equation}

We solve the Schr\"{o}dinger equation exactly by using the Floquet's theorem,\cite{shirley,cheng}
\begin{equation}
i\frac{\partial}{\partial t}\Psi({\bf r},t)=H({\bf k},t)\Psi({\bf r},t) \ .
\end{equation}
The solution is $\Psi_{s, {\bf k}}({\bf r},t)=\frac{1}{2\pi}e^{i{\bf k}\cdot{\bf r}}\Phi_{s,{\bf
  k}}(t)$. According to Ref. [\onlinecite{shirley}], $\Phi_{s,{\bf 
  k}}(t)=\phi_{s,{\bf k}}(t)e^{-iq_{s}({\bf k})t}$, where $q_{s}({\bf k})$ is the eigenvalue with $s=\pm$
  represent two branchs of soultion,  
  $\phi_{s,{\bf k}}(t)$ is a periodic function which satisfy
$[i\frac{\partial}{\partial t}+q_{s}({\bf k})]\phi_{s,{\bf k}}(t)=H({\bf k},t)\phi_{s,{\bf k}}(t)$.
Expand by Fourier series $\phi_{s,{\bf k}}(t)=\sum_{-\infty}^{\infty}\phi_{s,{\bf
    k}}^{n}e^{in\Omega t}$, $\phi_{s,{\bf k}}^{n}$ is a two component spinor $(\phi_{s,{\bf k}}^{n,A},\phi_{s,{\bf
    k}}^{n,B})^{T}$. Then, the eigenvalues and eigenfunctions for circular polarized field and linear 
  polarized field can be determined by
\begin{widetext}
\begin{eqnarray}
[n\Omega-q_{s}({\bf k})]\phi_{s,{\bf k}}^{n}+\tau \alpha[\sigma^{\tau}\phi_{s,{\bf k}}^{n+1}-\sigma^{-{\tau }}\phi_{s,{\bf
    k}}^{n-1}]+v_{F}(\sigma_{x}\tau k_{x}+\sigma_{y}k_{y})\phi_{s,{\bf k}}^{n}=0 \ ,\\
\label{circular}
[n\Omega-q_{s}({\bf k})]\phi_{s,{\bf k}}^{n}+\tau\alpha^{\prime}\sigma_{x}[\phi_{s,{\bf k}}^{n+1}-\phi_{s,{\bf
    k}}^{n-1}]+v_{F}(\sigma_{x}\tau k_{x}+\sigma_{y}k_{y})\phi_{s,{\bf k}}^{n}=0 \ .
\label{linear}
\end{eqnarray}
\end{widetext}
Here $\sigma^{\pm}=\sigma_{x}\pm i \sigma_{y}$, $\alpha=\frac{v_{F}eE_{0}i}{2\sqrt{2}\hbar\Omega}$, 
$\alpha^{\prime}=\frac{v_{F}eE_{0}}{2\hbar\Omega i}$.
It is obviouse that such equations have periodic structure with the integer number of $\Omega$. If we replace
$q_{s}({\bf k})$ by $q_{s}({\bf k})+l \omega$, $l$ is an arbitrary integer, one can find that the equations
are unchanged, both $q_{s}({\bf k})$ and $q_{s}({\bf k})+l \omega$
could be the eigenvalues. For convenience, we choose $-\Omega/2<q_{s}({\bf k})\leq /2$. 
Moreover, since $H({\bf k},t)$ is Hermitian, the eigenvalues must satisfy
$\sum_{s}q_{s}({\bf k})=\frac{1}{T_{0}}\int_{0}^{T_{0}}\mbox{Tr}H({\bf k},t)dt$,\cite{shirley} therefore we have $q_{-}({\bf
  k})=-q_{+}({\bf k})$ in our calculation.
The density of states (DOS) is given by
\begin{equation}
\rho(t_{1},t_{2})=\int d{\bf k} \sum_{s=\pm 1}\Phi_{s,{\bf k}}(t_{1})\Phi_{s,{\bf k}}^{\dag}(t_{2}) \ ,
\end{equation}
it is a $2 \times 2$ matrix in pseudospin space. 
In order to translate into energy space, one can let $T=t_{1}+t_{2}$ and $t=t_{1}-t_{2}$.\cite{haug} After
Fourier transformation with respect to $t$, one can obtain:
\begin{eqnarray}
\rho_{\xi_{1},\xi_{2}}(T,\omega)=&&\int d{\bf k}\sum_{s=\pm}\sum_{n.m=-\infty}^{\infty}R_{\xi_{1},\xi_{2}}(s;n,m;{\bf
  k})e^{i(n-m)\Omega T} \nonumber \\
&&\times \delta[\omega-(q_{s}({\bf k})-(n+m)\Omega/2)] \ ,
\label{dos}
\end{eqnarray}
where $\xi_{1}(\xi_{2})=A$ or $B$, and $R_{\xi_{1},\xi_{2}}(s;n,m;{\bf k})=\phi_{s,{\bf k}}^{n,\xi_{1}}(\phi_{s,{\bf
    k}}^{m,\xi_{2}})^{\dag}$. Then the electron density for pseudospin state $\xi$ is 
\begin{equation}
n_{\xi}=(1/2\pi)\int_{0}^{E_F(T)} d\omega \rho_{\xi,\xi}(T,\omega)\ .
\label{density}
\end{equation}
Once the total electron density $n^{K}=n^{K}_{A}+n^{K}_{B}$
($n^{K^{\prime}}=n^{K^{\prime}}_{A}+n^{K^{\prime}}_{B}$) in valley K (K$^{\prime}$) is given, one can determine the time-dependent Fermi energy
$E_{F}^{K}(T)$ ($E_{F}^{K^{\prime}}(T)$) according Eq.\ (\ref{density}).  
We define the pseudospin polarization in valley K (K$^{\prime}$) $P^{K}=P^{K}_{A}+P^{K}_{B}$ ($P^{K^{\prime}}=P^{K^{\prime}}_{A}+P^{K^{\prime}}_{B}$).
In order to avoid treating insignificant singularity in numerical calculation, we interchange the integration $d \omega$ and
$d{\bf k}$ in Eqs.\ (\ref{dos}) (\ref{density}), then we can obtain the relation between electron density and Fermi energy
without presenting DOS. 

\begin{figure}[htb]
\epsfig{figure=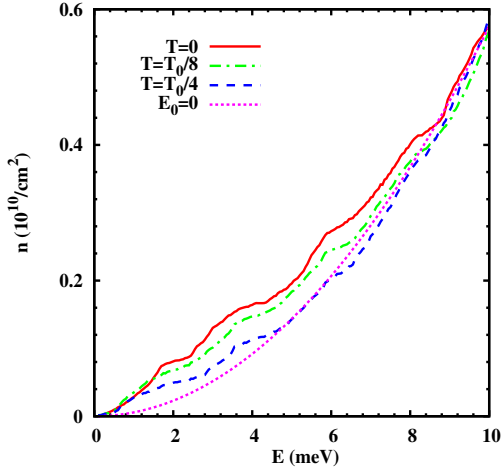,width=9cm,height=6.5cm,angle=0}
\vskip-0.3cm
\caption{(color online) Electron density versus Fermi energy in K (and K$^{\prime}$) valley under a linear
  polarized THz field with $E_{0}=1$ kV/cm and $\Omega=2\pi$ THz 
  at $T=0$ (solid curve), $T_{0}/8$ (dotted-dashed curve) and $T_{0}/4$ (dashed curve). They are in contrast with the
  dashed parabola when the THz field is absent (dotted curve).}
\label{fig1}
\end{figure}

First, in Fig.\  \ref{fig1}, we show the density-energy (n-E) curve whose differential with energy is DOS for linear polarized THz field at $T=0$,
$T_{0}/8$, $T_{0}/4$ . 
We also plot the parabolic n-E curve $n=\frac{1}{2\pi}(\frac{E_{F}}{\hbar v_{F}})^{2}$
without THz field.
It is obvious that n-E curve can be remarkably modified by THz field, and it (as well as DOS) vary with time as the
time-dependent electric field modulus. 
We pointed out that there is no pseudospin polarization in linear field case,
the reason the symmetry between sublattice A and B. If we interchange the index of A and B and let $k_{y}
\rightarrow -k_{y}$ in Eq.\ (\ref{linear}), the equation is unchanged. Since the summation over ${\bf k}$ in
Eq.\ (\ref{dos}), one can find $\rho_{A,A}(T, \omega) \equiv \rho_{B,B}(T,\omega)$.

\begin{figure}[htb]
\epsfig{figure=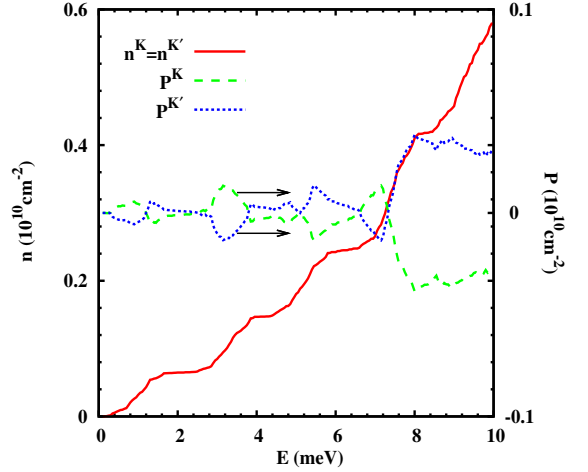,width=9cm,height=6.5cm,angle=0}
\vskip-0.3cm
\caption{(color online) Electron density and pseudospin polarization versus Fermi energy in K and K$^{\prime}$ valleys under
  a circular polarized THz field with $E_{0}=1$ kV/cm and $\Omega=2\pi$ THz. $n^{K}=n^{K^{\prime}}$ (solid curve) and
  $P^{K}$ (dashed curve) $=-P^{K^{\prime}}$ (dotted curve). All the curves do not vary with time.}
\label{fig2}
\end{figure}

The circular polarized THz field case is much more interest. We plot both the n-E curve and pseudospin
polarization-energy (P-E) curve of K and K$^{\prime}$ valleys in Fig.\ \ref{fig2}. 
Here, the DOS does not vary with time as the electric field modulus is a constant.
One can see that the n-E curves of two valleys are exactly identical and the P-E curves are exactly oppsite in
sign. This can be understood as the symmetry between $\tau= +$
and $-$ cases of Eq.\ (\ref{circular}). If we interchange the index of A and B for $\tau=-$ case, and then
let $q_{s}({\bf k})\rightarrow 2n\Omega+q_{s}({\bf k})$, the eqaution will be the same as $\tau= +$ case
besides $q_{s}({\bf k}) \rightarrow -q_{s}({\bf k})=q_{-s}({\bf k})$. Because of the summation over $s$ in
Eq.\ (\ref{dos}), we can easily find that $\rho_{A,A}^{K}(\omega)=\rho_{B,B}^{K^{\prime}}(\omega)$ and
$\rho_{B,B}^{K}(\omega)=\rho_{A,A}^{K^{\prime}}(\omega)$. However, the symmetry inside each valley we
discussed of linear field case in above paragraph no longer
come into existing as the appearence of $\sigma^{+}$ and $\sigma^{-}$ in Eq.\ (\ref{dos}), therefore
$\rho_{A,A}^{K}(\omega)\neq\rho_{B,B}^{K}(\omega)$ and
$\rho_{A,A}^{K^{\prime}}(\omega)\neq\rho_{B,B}^{K^{\prime}}(\omega)$. 
A direct result is that once the given electron densities of two
valleys that can be contraled by a gate voltage\cite{yan} are the same, in other words the Fermi energies of
two valleys are the same, then the total pseudospin polarization $P=P^{K}+P^{K^{\prime}}$ must be zero even
though $P^{K}\neq 0$ and $P^{K^{\prime}}\neq 0$. We point out that one can use the so-called ``valley filter
'' proposed by Recerz {\em et al.} in Ref.\ [\onlinecite{rycerz}] to make the electron density (and Fermi energy) of one valley be very
different from the other valley. Then, a total pseudospin polarization is obtained.

\begin{figure}[htb]
\epsfig{figure=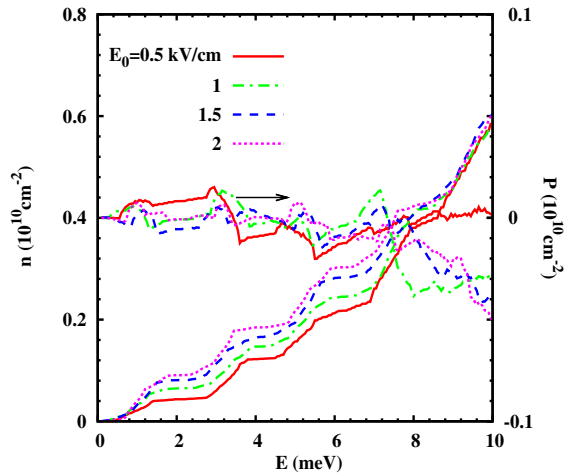,width=9cm,height=6.5cm,angle=0}
\vskip-0.3cm
\caption{(color online) Electron density and pseudospin polarization versus Fermi energy in K valley with
  $E_{0}=0.5$ (solid curve), $1$ (dotted-dashed curve), $1.5$ (dashed curve) and
  $2$ kV/cm (dotted curve), $\Omega=2\pi$ THz.}
\label{fig4}
\end{figure}

\begin{figure}[htb]
\epsfig{figure=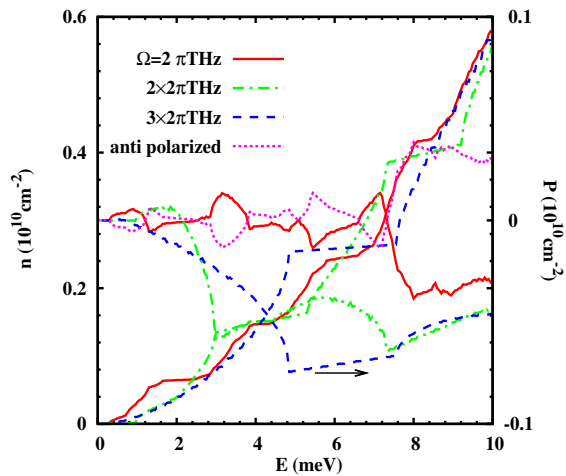,width=9cm,height=6.5cm,angle=0}
\vskip-0.3cm
\caption{(color online) Electron density and pseudospin polarization versus Fermi energy in K valley with  $E=1$ kV/cm and
  $\Omega=1$ (solid curve), $2$ (dotted-dashed curve),$3 \times2\pi$ (dashed curve)THz. An anti-orientated
  circular polarized fild case is also present (dotted curve).}
\label{fig5}
\end{figure}

Finally, we investigate the influence of the THz field parameters such as field strength, frequency and
polarization orientation in Fig.\ \ref{fig4} and Fig.\ \ref{fig5}. We find that the n-E and P-E curves strongly depend on the field
strength and frequency although. 
Especially, the polarization orientation can be utilized easily. The sign of pseudospin polarization can be changed if we
apply an antiorientated THz field eventhough the n-E relation is unchanged.
This feature may provide a very useful tool to manipulate the pseduspin polarization for graphene-based
devices in the future.

In Summary, we propose a method to generate and manipulate pseudospin polarization in a monolayer graphene
by using a circular polarized intense terahertz laser field. 
We solve the time-dependent Schr\"{o}dinger equation with help of Floquet's theorem in this paper. 
It is pointed out that the pseudospin polarization in K and K$^{\prime}$ valleys can be exactly opposite sign when the electron
densities in these two valleys are the same, and a total psedospin polarization can be obtained by utilizing
valley filter which makes densities to be different in two valleys.
Moreover, we find that the psedospin polarization can be manipulated by the strength,
frequency, and especially the polarization orientation of the field.

The author would like to thank M. W. Wu for proposing the topic as well as the directions during the investigation. 
The author would also like to thank J. H. Jiang, J. L. Cheng and Y. Zhou for helpful discussion.   
This work was supported by the 
Natural Science Foundation of China under Grant Nos.\ 10574120, 
the National Basic Research Program of China under Grant No.\ 2006CB922005 
and the Knowledge Innovation Project of Chinese Academy of Sciences.


\begin{thebibliography}{10}

\bibitem{novoselov1} K. S. Novoselov, A. K. Geim, S. V. Morozov, D. Jiang, Y. Zhang, S. V. Dubonos,
  I. V. Grigorieva, and A. A. Firsov, Science {\bf 306}, 666 (2004).

\bibitem{novoselov2} K. S. Novoselov, D. Jiang, F. Schedin, T. J. Booth, V. V. Khotkevich, S. V. Morozov, and
  A. K. Geim, Proc. Natl. Acad. Sci. USA, {\bf 102}, 10451 (2005).

\bibitem{novoselov3} K. S. Novoselov, A. K. Geim, S. V. Morozov, D. Jiang,M. I. Katsnelson,
  I. V. Grigorieva, S. V. Dubonos, and A. A. Firsov, Nature (London) {\bf 438}, 197 (2005).

\bibitem{geim1}A. K. Geim and K. S. Novoselov, Nat. Mater. {\bf 6}, 183 (2007)

\bibitem{geim2} A. K. Geim and A. H. MacDonald, Phys. Today {\bf 60} (8), 35 (2007).

\bibitem{zhang} Y. Zhang, Y. -W. Tan, H. L. Stormer, and P. Kim, Nature (London) {\bf 438}, 201 (2005).

\bibitem{zhou} S. Y. Zhou, G. -H. Gweon, A. V. Fedorov, P. N. First, W. A. de Heer, D. -H. Lee, F. Guinea,
  A. H. Castro Neto, and A. Lanzara, Nat. Mater. {\bf 6}, 770 (2007).

\bibitem{trau} B. Trauzettel, D. V. Bulaev, D. Loss, and G. Burkard, Nat. Phys. {\bf 3}, 192 (2007).

\bibitem{wallace}P. R. Wallace, Phys. Rev {\bf 71}, 622 (1947).

\bibitem{rycerz} A. Rycerz, J. Tworzydlo, and C. W. J. Beenakker, Nat. Phys. {\bf 3}, 172 (2007).

\bibitem{tan} S. G. Tan, M. B. A. Jalil, D. E. Koh, H. K. Lee, and Y. H. Wu, arXiv:0806.1568.

\bibitem{min} Hongki. Min, Giovanni Borghi, Marco Polini, and A. H. MacDonald, Phys. Rev. B {\bf 77}, 041407 (2008).

\bibitem{wolf} S. A. Wolf, D. D. Awschalom, R. A. Buhrman, J. M. Daughton, S. von Moln\'{a}r, M. L. Roukes,
  A. Y. Chtchelkannova, and D. M. Treger, Science {\bf 294}, 1488 (2001).

\bibitem{prinz}
{\em Semiconductor Spintronics and Quantum
  Computation}, eds. D. D. Awschalom, D. Loss, and N. Samarth
  (Springer, Berlin, 2002); I. \v Zuti\'c, J. Fabian, and S. Das Sarma,
Rev. Mod. Phys. {\bf 76}, 323 (2004).

\bibitem{johnsen} K. Johnsen, Phys. Rev. B {\bf 69}, 10978 (2000).

\bibitem{cheng} J. L. Cheng and M. W. Wu, Appl. Phys. Lett. {\bf 86}, 032107 (2005)

\bibitem{jiang1} J. H. Jiang, M. Q. Weng, and M. W. Wu, J. Appl. Phys. {\bf 100}, 063709 (2006); Y. Zhou,
  Physica E {\bf 40}, 2847 (2008).

\bibitem{jiang2} J. H. Jiang and M. W. Wu, Phys. Rev. B {\bf 75}, 035307 (2007); J. H. Jiang, M. W. Wu, and Y. Zhou, arXiv:0805.3280.
 
\bibitem{shirley} J. H. Shirley, Phys. Rev. {\bf 138}, B979 (1965).

\bibitem{haug} H. Haug and A. P. Jauho, {\em Quantum Kinetics in
Transport and Optics of Semiconductor} (Spinger-Verlag, Berlin, 1996).

\bibitem{yan} J .Yan, Y. B. Zhang, P. Kim, and A. Pinczuk, Phys. Rev. Lett. {\bf 98}, 166802 (2007).

\bibitem{kane} C. L. Kane and E. J. Mele, Phys. Rev. Lett. {\bf 95}. 226801 (2005).
\end{thebibliography}
\end{document}